\nonstopmode
\documentclass[prc,amsmath,showpacs,twocolumn]{revtex4}
\begin{document}
~\\\begin{flushright}TRIUMF preprint: TRI-PP-05-06\end{flushright}

\title{The Shell Model, the Renormalization Group and the Two-Body Interaction}
\author{B.K.~Jennings}\email{jennings@triumf.ca} \affiliation{TRIUMF, 4004
Wesbrook Mall, Vancouver, BC, Canada, V6T 2A3} \date{\today}
\begin{abstract}
The no-core shell model and the effective interaction $V_{{\rm low}\ k}$ can
both be derived using the Lee-Suzuki projection operator formalism. The main
difference between the two is the choice of basis states that define the model
space. The effective interaction $V_{{\rm low}\ k}$ can also be derived using
the renormalization group. That renormalization group derivation can be
extended in a straight forward manner to also include the no-core shell
model. In the nuclear matter limit the no-core shell model effective
interaction in the two-body approximation reduces identically to $V_{{\rm low}\
k}$. The same considerations apply to the Bloch-Horowitz version of the shell
model and the renormalization group treatment of two-body scattering by Birse,
McGovern and Richardson.
\end{abstract}
\maketitle

{\it Ab initio} shell model calculations are becoming a staple of nuclear
structure calculations\cite{haxtona,noc,barrett,zhan}. The motivation is to
start with a bare two-nucleon potential or a bare two-nucleon plus bare
three-nucleon potential and calculate nuclear properties from that potential
making only controllable approximations. The potential is taken as given. A
parallel development\cite{birse,schwenk,vlk} is taking place in the study of
the nucleon-nucleon potential where effective interactions are being developed
through the use of renormalization group techniques. Here the motivation is not
just to study nuclear structure but also to understand the nature of the
two-body and many-body forces. It is the purpose of the present paper to show
that these process should be considered convergent rather than parallel. The
main and indeed the only significant difference between the two approaches is
the choice of the basis that define the projection operator and hence the model
space. The shell model uses the harmonic oscillator basis while the two-body
work uses a plane wave basis.

To be more definite, first consider the no-core shell model. It uses the
Lee-Suzuki projection formalism\cite{lees,lee}. A word is needed about what
exactly is meant by the Lee-Suzuki formalism since there are variants that
could be confusing if not kept straight. The key aspect\cite{previous} of the
Lee-Suzuki method is the determination of the the operator $\omega$ defined by
the equation:
\begin{eqnarray}
Q\omega P |\psi\rangle = Q|\psi\rangle
\end{eqnarray}
where $P$ and $Q=1-P$ are the projection operators that define the model
space. In the Lee-Suzuki formalism we define a set of states $|k\rangle$ that
are mapped into the model space defined by $P$. In terms of the states
$|k\rangle$ and a complete orthonormal set of states $|\alpha_P\rangle$ that
span the model space we have:
\begin{eqnarray}
Q\omega P |\alpha_P\rangle = \sum_k Q|k\rangle \langle \alpha_P | k
\rangle^{-1}
\end{eqnarray}
Once $\omega$ is determined we can calculate\cite{previous} the effective
interaction. We can have either a non-hermitian version of the interaction,
which is simpler, or an hermitian version which is more complicated. For the
present purposes the non-hermitian version is preferred.

In the no-core shell the projection operator, $P$, is defined in terms of
harmonic oscillator states. For a given oscillator parameter, $\hbar\Omega$,
the states up to a given energy are used. In addition an harmonic oscillator
term in the center of mass is added. While this latter term is very important
in the numerical work it plays no role in our discussion.

In the derivation of $V_{{\rm low}\ k}$\cite{schwenk} the Lee-Suzuki formalism
is also used but with the projection operator define by plane wave states. All
plane wave states below a given momentum are included in the model space. Since
$V_{{\rm low}\ k}$ can also be derived in a renormalization group approach it
is normal to ask if, in general, the Lee-Suzuki formalism can be cast into a
renormalization language. For a class of projection operators, including the
one used in the no-core shell model, this is indeed possible. Once the
formalism is set up it is trivial. The main difference with the derivation of
$V_{{\rm low}\ k}$ is that we have discrete states.

The procedure is to recast the discrete state problem into a from that
resembles the Lippmann-Schwinger equation. Then either $V_{{\rm low}\ k}$ or
Birse formalism\cite{birse} can be applied line for line to get a flow equation
for the shell-model interaction. We define a Hamiltonian $H=H_o+\delta H$ and
an unperturbed Hamiltonian $H_o$ that is used to define the projection operator
$P=\theta(\Lambda-H_0)=\sum_n^N|\psi^n_o\rangle\langle\psi^n_o|$. Here $N$
denotes the highest energy state in the model space and the eigenvalue
equations are given by:
\begin{eqnarray}
H_o|\psi^n_o\rangle&=&E^n_o|\psi^n_o\rangle\\
H|\psi^n\rangle &=& E^n|\psi^n\rangle
\end{eqnarray}
Next a t-matrix like quantity is defined by:
\begin{eqnarray}
T|\psi^n_o\rangle = \delta H |\psi^n\rangle
\end{eqnarray}
As in scattering theory (which this is not) we have the following equations:
\begin{eqnarray}
|\psi^n\rangle &=& |\psi^n_o\rangle + G_o(E) \delta H |\psi^n\rangle\\
               &=&  |\psi^n_o\rangle + G_o(E) T |\psi^n_o\rangle\\
T&=&\delta H + \delta H G_o(E) T
\end{eqnarray}
Next we define the matrix elements $T^{n',n}=\langle\psi^{n'}_o | T |
 \psi^n_o\rangle$ and $V^{n',n}=\langle\psi^{n'}_o | \delta H |
 \psi^n_o\rangle$. The symbol $V$ is chosen to make the results look more
 familiar. The Lippmann-Schwinger equation is now written:
\begin{eqnarray}
T^{n',n}=V^{n',n}+\sum_{n''}V^{n',n''} \frac{1}{E^n-E^{n''}_o} T^{n'',n}
\end{eqnarray}
The expansion coefficients defined by $|\psi^n\rangle = \sum_{n'} a^n_{n'}
|\psi^{n'}_o\rangle $ are given by:
\begin{eqnarray}
 a^n_{n'} = \delta^{n',n} + \frac{1}{E^n-E^{n'}_o} T^{n',n}
\end{eqnarray}
Note that there are no sums in this last equation. 

The effect of the projection is to restrict the intermediate sums. We can now
define an effective interaction by:
\begin{eqnarray}
T^{n',n}=V_N^{n',n}+\sum_{n''}^N V_N^{n',n''} \frac{1}{E^n-E^{n''}_o} T^{n'',n}
\end{eqnarray}
where the effective interaction, $V_N^{n',n}$, is chosen to keep the half-off
shell matrix $T$ or the fully off shell $T$ independent of the cutoff. It is
here and in similar sums that the explicit form of the projection operator
given above is needed.

We can now do two different developments. First we can follow the proponents of
$V_{{\rm low}\ k}$ and work with the half-off shell version of the
$T$-matrix. The procedure is exactly as they did\cite{schwenk} with the
integrals replaced by sums. The resulting flow equation is:
\begin{eqnarray}
V_{N\ LS}^{n',n} - V_{N-1\ LS}^{n',n} = V_{N\ LS}^{n',N}\label{eq_ls}
\frac{1}{E^N_o -E^n_o} T^{N,n}
\end{eqnarray}  
Note that it is the unperturbed energies that occur in this expression. By the
arguments in ref.~\cite{previous} this corresponds to the Lee-Suzuki projection
formalism (hence the LS subscript). We stress again that it is the
non-hermitian version (eq.~3 of ref.\cite{previous}) of the Lee-Suzuki
interaction that agrees with the renormalization group interaction.

Second we can follow Birse {\em et al} \cite{birse} and use the fully off shell
$T$-matrix to recover\cite{previous} the Bloch-Horowitz projection formalism. A
less general derivation is given in ref.~\cite{haxton}.  Again the only
difference is that the integrals are replaced by sums. The resulting flow
equation is:
\begin{eqnarray}
V_{N\ BH}^{n',n} - V_{N-1\ BH}^{n',n} = V_{N\ BH}^{n',N}
\frac{1}{E^N_o -E} V_{{\cal N}\ BH}^{N,n}\label{eq_bh}
\end{eqnarray}
where $E$ is the value of the energy where the $T$ matrix is required. This
equation is valid for $n\le N-1$. The value of $\cal N$ is not uniquely
determined but rather the equation is valid for any ${\cal N} \ge N-1$. For
many purposes the value ${\cal N}=N$ is the most useful.

The difference between eqs.~\ref{eq_ls} and \ref{eq_bh} is the usual trade off
between $G_o T$ and $G V$ and indeed the derivation of $V_{{\rm low}\ k}$ uses
that equivalence.  Eq.~\ref{eq_ls} requires that $T$ be calculated at each
step. The result is energy independent.  On the other hand, eq.~\ref{eq_bh}
just uses $V$ but must be calculated separately for each energy.  Although the
two effective potentials are related\cite{bogner} they have different
properties\cite{previous} and are not equal.  Since potentials are not
observables, there is no particular reason they should be.
 
Since both the no-core shell model and the $V_{{\rm low}\ k}$ interaction can
be derived from the renormalization group we see that using $V_{{\rm low}\ k}$
in a no-core shell-model calculation may not be very beneficial. Certainly
there would be little gained in re- renormalizing an interaction that has
already been renormalized. Using $V_{{\rm low}\ k}$ directly in the oscillator
diagonalization would probably be preferable as it already has suppressed the
high momentum components.

The essential difference between shell models (either Lee-Suzuki or
Bloch-Horowitz) and the corresponding continuum two-body calculations, $V_{{\rm
low}\ K}$ or Birse {\em et al}\cite{birse}, is the choice of projection
operator: harmonic oscillator vs plane wave. However there is one limit in
which the two become the same: namely the nuclear matter limit. This limit
consists of letting the $\hbar \Omega \rightarrow 0$, $N\rightarrow \infty$
while keeping the cutoff energy, $E_{\rm cutoff}$ fixed (in one dimension
$(N+1/2) \hbar \Omega=E_{\rm cutoff}$). This is easiest to see in one dimension
where the harmonic oscillator wave functions are a Gaussian times an Hermite
polynomial. The asymptotic form for of the Hermite polynomial is given by (see
ref.~\cite{grad} eqs. 8.955) :
\begin{eqnarray}
H_{2n}(x)=\frac{(-1)^n 2^{2n}}{\sqrt{\pi}}e^{x^2/2} \Gamma(n+1/2)\nonumber\\
  \cos(\sqrt{4n+1}x)
\end{eqnarray}
with a similar equation for the odd Hermite polynomials in terms of sine
functions. The argument of the cosine is just $r\sqrt{ 2 m E}/\hbar =rp/\hbar $
once the oscillator length parameter is inserted and $n \hbar \Omega$ is
expressed in terms of $E$. The coefficient of the cosine cancels the oscillator
normalization to within a factor of $\sqrt\pi /2 $ when Sterling's formula
(ref.~\cite{grad} eq.~8.327) is used for the gamma functions. Since we are
interested in large $n$ this is valid.

Thus we see that in the nuclear matter limit the harmonic-oscillator
shell-model projection operator reduces to the momentum space projection
operator. If the shell model is carried out to the two-body cluster
approximation then the effective interaction reduces identically to $V_{{\rm
low}\ k}$ (Lee-Suzuki) or Birse {\em et al}\cite{birse} (Bloch-Horowitz).
Therefore nuclear matter, studies such as ref.~\cite{furn}, are relevant to the
no-core shell model even when $V_{{\rm low}\ k}$ is not explicitly used in the
shell model calculation. Or even more interestingly refs.~\cite{haxtona} and
ref.~\cite{furn} are partially discussing the same thing and coming to the same
conclusion: namely that when done properly nuclear physics is perturbative.
They did this although one was using Bloch-Horowitz in finite nuclei and the
other was using Lee-Suzuki in nuclear matter.

Even more information relevant to the shell model and nuclear physics in
general can be extracted from the nuclear matter results. Ref.~\cite{furn},
working in nuclear matter, emphatically makes the point that a higher cutoff
(momentum or energy) is not always better. Applying the same reasoning to
finite nuclei we infer that shell-model calculations may not gain by taking
larger and larger model spaces. But rather the model space should be matched to
the energy and momentum scales of the problem since making the space larger may
just induce spurious loop contributions that then have to be canceled, for
example, by many-body forces. Ref.~\cite{nogga} (especially fig.~1) indicates
that higher cutoffs require stronger many-body forces. Indeed nuclear physics
has been plagued by the need to cancel loop contributions. The success of Dirac
phenomenology\cite{serot} has been traced to the fact that it does a better job
of suppressing\cite{thies,cooper} high loop momenta than non-relativistic
calculations. In pion-nucleus scattering there is similar\cite{thies} need to
suppress\cite{rostokin,detak} high momentum loop contributions. In the later
case this is frequently discussed under that name of the EELL
effect\cite{eell}.

Interactions with low cutoffs, in either momentum or oscillator space, cannot
be considered more or less fundamental than those with high cutoffs. This is
especially true for those obtained in the Lee-Suzuki formalism since a unitary
transformation underlies\cite{lee} this approach. Unitary transformations
change the appearances but not the predictions or underlying physics. Thus one
can not say $V_{{\rm low}\ k}$ is any more or less fundamental than the higher
cutoff potentials it is related to. The arguments apply equally well to the
shell model effective interactions. Does the use of an harmonic oscillator
basis make them less fundamental than those in a plane wave basis? Thus we can
consider the no-core shell-model model-space interactions just as fundamental
as $V_{{\rm low}\ k}$. In nuclear matter they become the same anyway. This
discussion is not meant as a criticism of $V_{{\rm low}\ k}$ but rather meant
to suggest a different way of looking at both $V_{{\rm low}\ k}$ and the shell
model effective interaction.

We noted at the beginning the rather different motivations for the shell model
and two-body continuum calculations: the difference between using and
generating an interaction. The discussion in this paper suggests the
distinction is not all that clear cut. The interaction must be matched to the
model space. Even the concept of a fundamental interaction is fraught with
difficulties.

\noindent{\bf Acknowledgment}

This work was motivated by discussions with A.~Schwenk. The Natural Sciences
and Engineering Research Council of Canada is thanked for financial support.

\end{document}